# The Diversity of Exoplanets: From Interior Dynamics to Surface Expressions

Maxim D. Ballmer[1,2] and Lena Noack[3]

## ABSTRACT

The coupled interior–atmosphere system of terrestrial exoplanets remains poorly understood. Exoplanets show a wide variety of sizes, densities, surface temperatures, and interior structures, with important knock-on effects for this coupled system. Many exoplanets are predicted to have a "stagnant lid" at the surface, with a rigid stationary crust, sluggish mantle convection, and only minor volcanism. However, if exoplanets have Earth-like plate tectonics, which involves several discrete, slowly moving plates and vigorous tectono-magmatic activity, then this may be critical for planetary habitability and have implications for the development (and evolution) of life in the galaxy. Here, we summarize our current knowledge of coupled planetary dynamics in the context of exoplanet diversity.

## INTRODUCTION

In the last 25 years, >4,000 of exoplanets have been discovered, and this progress is expected to continue with ongoing and planned missions such as the *PLATO* (*PLAnetary Transits and Oscillations of stars*) space telescope. The exoplanets found so far display a wide range of masses and sizes, pointing to a diversity in compositions and structures. Many planets are sufficiently small and have a density that implies mostly rocky compositions, i.e., with a silicate envelope (or "mantle") around a metallic core. This fundamental observation demonstrates that planets broadly similar to Earth (or Venus or Mars) are common in the galactic neighborhood.

However, the architectures of exoplanetary systems are often surprising. For example, initial discoveries revealed giant planets with very close orbits around their host stars. While these "hot Jupiters" turned out to be exceptions, the most commonly observed types of exoplanets also do not have any analogy in our own Solar System: a "typical" exoplanet appears to be a mostly rocky planet with a mass several times that of Earth and with a significant envelope of steam or ice (referred to as "sub-Neptunes"), or without such a hydrous envelope ("super-Earths") (Fulton et al. 2017). Even among the discovered exoplanets with roughly Earth-like properties in terms of mass and radius, a large variety persists in terms of major-element composition, atmospheric properties, or equilibrium surface temperature (as inferred from the distance to its host star).

[1] University College London
Department of Earth Sciences
E-mail: m.ballmer@ucl.ac.uk

[2] Tokyo Institute of Technology
Earth-Life Science Institute
E-mail: ballmer@elsi.jp

[3] Freie Universität Berlin
Institute of Geological Sciences
E-mail: lena.noack@fu-berlin.de

The discovery of exoplanets has intrigued specialists and laypersons alike. This excitement is boosted by the possibility of extraterrestrial life, with implications for the genesis of life on Earth, and, thus, our own roots. One requisite for life (as we know it) is the presence of liquid water on the surface of a planet, a condition potentially met in the "habitable" zone around a star (Kasting et al. 1993). With new observations, rocky exoplanets with similar masses and radii as the Earth may be discovered within the "habitable zone" of a Sun-like star. Whether such planets can sustain liquid water and life-friendly surficial conditions, however, depends not only on the star–planet distance and host-star temperature but also on the planet's interior dynamics and tectonic regime (Dehant et al. 2019).

Earth is unique in our Solar System in that it hosts life and oceans, as well as displaying plate tectonic activity. Today's Earth surface is divided into ~10 large plates, and several smaller ones, which move with typical speeds of a few centimeters per year. Plate tectonics describes and explains plate motions as they spread apart to form new sea floor; slide past one another, as at the San Andreas Fault (California, USA, extending into Mexico); or collide, with one plate being subducted beneath the other. Most of terrestrial volcanism and crustal deformation (with related earthquakes) occurs along plate boundaries. How and when plate tectonics started, however, remains controversial: it has likely been active for at least half, and perhaps up to two thirds (or even more) of Earth's 4.5 Gy history (Korenaga 2013). In contrast, most of the other rocky planets in our Solar System display little or no surficial evidence of tectonic activity. They are locked in a "stagnant-lid" regime, having a thick lithosphere that does not significantly deform or move laterally (FIG. 1). The only volcanic activity on stagnant-lid planets is localized above deep-seated mantle upwellings ("plumes"), where heat can penetrate the lithosphere. So, while on Earth, oceanic lithosphere is recycled on time-scales of <100 million years, the lithospheres on Mars, the Moon, and Mercury are billions of years old, as evidenced by their heavily cratered surfaces.

In recent decades, it has become more and more established that plate tectonics may be key to sustaining life-friendly conditions (Ward and Brownlee 2003). On Earth, the subduction of carbonates ($CO_2$-bearing rocks) into the mantle is closely linked to climate evolution. Major volcanic activity occurs at plate boundaries, replenishing volatiles into the atmosphere as well as nutrients in the soil. Plate tectonics may also be essential to expose diverse rock types at the surface and form diverse landscapes, regulate ocean mass, and promote the evolution of life. Ironically, some of the major natural hazards such as earthquakes, the often-related tsunamis, or volcanic eruptions are the surface expression of the same processes that sustain life-friendly conditions: mantle convection and plate tectonics.

If plate tectonics is indeed critical for protracted biological evolution, a large fraction of exoplanets, even if located in the habitable zone, may not host complex lifeforms. Of course, the likelihood of life in the galactic neighborhood is extremely difficult to quantify. So far, we know of only one planet that hosts life, and we have very little knowledge of the potentially variable conditions that can lead to carbon-based life, or even "life" that is based on alternative—and yet unknown—biochemical cycles. At this stage, we cannot even exclude that basic life exists on other planetary bodies in our own Solar System, e.g., potentially in the higher clouds of the otherwise hostile environment on Venus (Greaves et al. 2020). This article does not set out to quantify the probability of life in the galaxy, but merely to juxtapose the processes that sustain life-friendly conditions on Earth with the variety of exoplanetary conditions.

## THE EARTH SYSTEM

All terrestrial planets, including Earth, likely start off hot due to massive energy release during accretion and core formation. Potential energy is released due to ever larger planetoids impacting the growing planet and due to the segregation of dense iron from rock to form the mantle and core. This energy is sufficient to promote largescale melting of most of the rocky mantle, referred to as the early "magma ocean" stage. As the magma ocean cools and partially solidifies, a primary atmosphere is outgassed and, probably within a few million years after accretion, a solid rocky mantle and crust are stabilized. At this stage, the planet is still hot and being sustained by the decay of radioactive nuclides. This primordial and radiogenic heat drives mantle convection over billions of years. The solid rocky mantle can flow at rates of centimeters per year, sustaining convective heat transport to the surface. For planets with a "mobile lid" (e.g., where plate tectonics is active), mantle cooling is very efficient, as the oceanic plates, which have cooled at the surface for tens to hundreds of millions of years, are subducted deep into the mantle. For stagnant-lid planets, the cooling of the mantle, and hence of the entire planet, is much less efficient: heat escape to the surface needs to be accommodated through the thick and stiff lithospheric lid (FIG. 1).

As the mantle cools efficiently on mobile-lid planets, the metallic core is also cooled efficiently. The cooling of the core–mantle boundary sets up a thermal gradient across the core, thereby driving vigorous thermal convection in the liquid outer core. Such convective flows stabilize the Earth's magnetic field in a process called the "geodynamo" (Roberts and Glatzmaier 2000). As in many technical applications, the motion of charged matter (in this case, liquid metal) induces an electromagnetic field. The Earth's magnetic field protects, in particular, land-based higher life forms from highly energetic solar particles. As the cooling of the mantle in stagnant-lid planets is inefficient, they usually cannot stabilize a planetary dynamo.

Most importantly, the cycling of materials (rocks and volatiles) and related chemical evolution of the Earth system is controlled by its plate tectonic surface dynamics (FIGS. 1 and 2). Materials are continuously introduced into the mantle at subduction zones, where they are stirred over timescales of hundreds of million years, and eventually recycled at mid-ocean ridges or other sites of volcanism, such as intraplate hotspots [e.g., Réunion (France) or Hawai'i (USA)], where hot mantle upwellings reach the base of the lithosphere. In the context of planetary habitability, this material cycling is most relevant for the major greenhouse gases of $H_2O$ and $CO_2$, as well as essential biological nutrients. Water can enter the mantle as subducted hydrous sediments or as hydrated oceanic crust. It is fixed, usually as $(OH)^-$, in the atomic structure of hydrous minerals, which are created when water circulates through the seafloor. Much of this mineral-bound water is released before it reaches the hot depths of the mantle, and is thought to promote melting and volcanism above subduction zones (the short-term volatile cycle). Nevertheless, a significant amount of the water and $CO_2$ remains fixed within the minerals that formed in the coolest, deepest parts of the subducted oceanic lithosphere. These fixed volatiles can reach the deep mantle (the long-term global volatile cycle) (Rüpke et al. 2004).

This efficient cycling and recycling of volatiles due to plate tectonics is the most important process to stabilize Earth's atmospheric composition and climate over millions, and billions, of years (FIG. 2). Although the outgassing of $CO_2$ and water from the mantle due to volcanism, as well as the ingassing into the mantle due to subduction, have varied through time, the surface volatile budget has remained largely balanced. For example, and as suggested by the available geological evidence, the volume of the oceans has not significantly varied over at least the last several hundred million years (Haq and Schutter 2008). A quasi-equilibrated budget of surface volatiles is readily reached as the huge mantle reservoir, through which the cycling occurs, releases and absorbs C and H at similar rates. Although estimates vary, the mantle and crust likely host several times as much $H_2O$ and >100,000 times as much carbon as the atmosphere, biosphere, and oceans combined. In contrast, the global volatile cycle on stagnant-lid planets mostly resembles a one-way street. While outgassing continuously occurs due to localized hotspot volcanism, ingassing remains very inefficient due to the lack of subduction. Therefore, the surface volatile budget does not reach a steady state until the mantle is nearly completely outgassed. In other words, the bulk volatile budget of a stagnant-lid planet needs to be "just right" to allow for long-term habitable conditions. If the total $CO_2$ budget is too large, a stagnant-lid planet may evolve into a hot "runaway greenhouse" state with atmospheric temperatures exceeding the boiling point of water and so potentially sterilizing the entire planet.

Another key process to buffer surface temperatures and climate is silicate weathering (FIG. 2). This process refers to the erosion of rocks and weathering of the (silicate) minerals near the surface. Silicate weathering draws greenhouse gases $CO_2$ and $H_2O$ out of the atmosphere, fixing them as carbonates and clay minerals. Since erosion and weathering are more efficient with higher surface temperatures (more precipitation, and also more $CO_2$ dissolved in the rain), the related process acts as a so-called negative feedback loop: with higher temperatures and stronger weathering, more $CO_2$ is drawn out of the atmosphere, buried as carbonates in the oceans and eventually subducted into the mantle, which ultimately decreases the temperature again. In turn, for cool temperatures, such as for a planet largely covered by ice in the extreme case (Kirschvink 1992), volatile release due to volcanism exceeds drawdown due to silicate weathering such that temperatures increase again. This important negative feedback loop operates on short geologic timescales of about one million years, efficiently stabilizing the climate. However, silicate weathering depends on many factors, such as land exposure, and is much more efficient on plate-tectonic planets such as Earth than on stagnant-lid planets with minor supply of fresh rocks. On plate-tectonic planets, volcanism is much more massive than on stagnant-lid planets, and mountain ranges are steadily built where two plates crash into another, both supplying fresh rock for erosion (West et al. 2005). By contrast, weathering on Venus and Mars is limited due to the low volumes of fresh volcanic rock and a lack of liquid surface water, such that the atmosphere is dominated by $CO_2$. Even for a stagnant-lid planet with abundant liquid surface water, weathering readily becomes supply limited, and the negative feedback loop is interrupted: any $CO_2$ release (e.g., due to volcanism) leads to global warming, $H_2O$ evaporation, and further related warming ($H_2O$ is an efficient greenhouse gas). This vicious cycle has been termed the "runaway greenhouse effect" (Ingersoll 1969) and has very likely occurred on Venus, a planet that is otherwise very similar to Earth (e.g., in terms of mass, radius, and bulk composition). The rather small difference between both planets in terms of distance to the Sun has apparently been sufficient to push Venus, but not the Earth, into a runaway greenhouse (Hamano et al. 2013). At least today, plate tectonics is prevented on Venus because of its high surface temperatures (also see below). Because stagnant-lid planets have a small supply of fresh rock for weathering, they must also have very small total volatile budgets to avoid a runaway greenhouse effect and be potentially habitable (Foley and Smye 2018).

## THE VARIETY OF EXOPLANET ENVIRONMENTS

As we focus beyond the Solar System to exoplanets, observational data are decidedly limited. First-order questions include: do these planets have plate tectonics and/or continental crusts? Are they volcanically active? Do they sustain habitable surface conditions? What about their long-term evolution as a coupled planetary system? At this stage, we need to rely on physical, chemical, and computational models to predict the possible interior and surface states of any given exoplanet, and infer their long-term evolution from the sparse data that we can gather from observations, stellar evolution models, and statistical data.

The critical importance of plate tectonics, in terms of sustaining life-friendly conditions, raises the question of its occurrence on other rocky planets beyond the Solar System. For most exoplanets, we only know the planet's mass and/or radius. Both quantities combined give the planet's mean density, which already provides a first indication towards its possible composition. A high density indicates a large quantity of heavier elements in the planet—Mercury's high density is related to its immense iron core making up about 80% of the planet's radius. A very low density, below what we would expect for an Earth-like planet, might indicate either a low metal content or a large fraction of light material, such as a thick $H_2O$ envelope.

Planets potentially covered with a deep ocean ("aquaplanets") have raised many speculations on the possible habitability of such bodies. For very deep $H_2O$ layers, high-pressure ices can form at the bottom of such oceans, as inferred for Jupiter's moon Ganymede. Such a high-pressure ice layer may chemically insulate the mantle (with its nutrients, salts, and carbon-storage facilities) from the water ocean, although some material exchange between the mantle and surface may still be possible (Noack et al. 2016; Kalousová et al. 2018). The deeper the water layer (and the thicker the possible underlying ice layer) the higher are the pressures acting on the rocky part of the planet. These high pressures may shut off melting, and, hence, interrupt global volatile cycles. This would demand that higher mantle convective stresses would be needed to break the lithosphere into a mosaic of plates and sustain plate tectonics (Kite et al. 2009; Noack et al. 2016). In turn, for massive super-Earths that do not maintain a sufficiently dense atmosphere after accretion, major volcanic activity, such as due to plate tectonics, may be essential to regulate surface temperatures and to allow for the presence of liquid water. With a stagnant lid, volcanic activity is much decreased (FIG. 1), particularly for massive planets, which may then not be able to outgas sufficient greenhouse gases to sustain a moderate climate (Dorn et al. 2018).

Because planet mass and size are the main observables of exoplanets, a rather strong debate has emerged around the question of whether these observables play a major role in sustaining a plate-tectonic or stagnant-lid regime. Several modeling studies have been conducted with apparently conflicting results (FIG. 3). Interestingly, very different trends in predicting the likelihood of plate tectonics as a function of planet mass can be explained when investigating the parameter spaces and approaches used in these studies (Stamenković and Breuer 2014). In general, the likelihood of plate tectonics is connected to mantle viscosity. A planetary mantle with low viscosity will convect vigorously, with narrow up- and downwellings, but this system can only sustain very low stresses, leading to a decoupling of the deep convecting part of the mantle and the cold and rigid surface layer (the lithosphere), which then evolves into a stagnant lid. As the mantle becomes more viscous, broad mantle upwellings are stabilized that dominate mantle mixing, leading to high stresses that then act on the lithosphere, localizing deformation and sustaining the formation of separate plates (van Heck and Tackley 2008). Once a mosaic of plates has formed, the stresses required to sustain mid-ocean ridge spreading and subduction (along with related volcanism and mountain building) (FIG. 1) are much smaller than those that are needed to break the plates, and are, therefore, less dependent on mantle viscosity. On Earth, these stresses are mostly accommodated by the weight of the subducted slabs that sink into the mantle (Forsyth and Uyeda 1975). However, for even more viscous mantles, mantle convection becomes very sluggish and convective stresses decrease again, such that the initiation of plate tectonics in the first place becomes less and less likely. Models therefore predict that the viscosity in the mantle should neither be too high nor too low in order to initiate and sustain plate tectonics. The likelihood of plate tectonics on an exoplanet depends on how the mantle viscosity profile changes as a function of planet mass or other factors (FIG. 3). Mantle viscosity itself mainly depends on the temperature of the mantle, and, hence, also on the budget of radioactive heat sources, which varies between planets. It is also sensitive to volatile content such as $H_2O$, which tends to weaken and lubricate minerals upon incorporation into their atomic structure. Viscosity generally increases with increasing pressure, such as in the deep interior of super-Earths, but the mineral structures at very high pressures and their potential to incorporate volatiles remain poorly understood.

In addition to mantle viscosity, the planetary surface temperature also influences the initiation of plate tectonics. Indeed, surface temperature is a potential observable through measurements of stellar flux and detections of greenhouse gases in exoplanetary atmospheres (Schaefer 2021, this issue). At very low surface temperatures, a thick stagnant lithospheric lid is expected to form, and the initiation of plate tectonics is unlikely. At moderate surface temperatures (e.g., similar to that of Earth), stresses at the base of the (thinner) lithosphere can become just high enough to sustain plate tectonics as long as other conditions are also favorable. For higher temperatures, plates boundaries become less well defined, but the overall behavior may still resemble plate tectonics, at least periodically, as has been suggested for the early Earth several billion years ago and present-day Venus (Davaille et al. 2017; Lourenço et al. 2020). Surface temperatures of several hundred degrees Celsius might lead to a hot and very weak lithospheric lid, critically reducing the stresses that are imposed by convection such that the initiation of plate tectonics becomes difficult, or that mobile-lid behavior occurs only intermittently ("episodic lid"), as has been also suggested for Venus (Landuyt and Bercovici 2009). Such very hot planets may further display voluminous volcanism, where lavas are continuously stacked upon each other such as on Jupiter's moon Io, a process called "heat piping" that may also play a role in preventing plate tectonics (Moore et al. 2017). For even higher surface temperatures, as may occur when a planet is in very close proximity to its host star, the (shallow) mantle and lithosphere may be fully molten to sustain a long-lived magma ocean with turbulent convection that incorporates the surface, but without relevance for plate tectonics, let alone habitability. Better constraints on the exact style of surface deformation on Earth several billion years ago, when the first bacterial life appeared, may provide important clues on the required conditions for the initiation of plate tectonics, and/or the origin of life.

Another factor for the liklihood of plate tectonics on a rocky planet is the planet's interior composition and structure. The chemistry of the building blocks of planets (especially in terms of major rock-forming cations Si, Mg, and Fe) are thought to be closely related to the composition of the host star, which can be observed through spectroscopic observations of its atmosphere (Putirka et al. 2021, this issue). Accordingly, the interior structure of a planet can be estimated from stellar composition, even though the processes that govern planetary formation from the stellar nebula add a significant degree of uncertainty. Such predicted interior structures help to bracket the possible range of expected metal core sizes and mantle compositions. Numerical models indicate that the patterns and vigor of mantle convection change with core size and directly influence the likelihood of plate tectonics, with an apparent optimum for relative core sizes slightly above that of Earth (Noack et al. 2014). The ratio of major elements Mg/Si (and also Fe/Mg) controls the viscosity of mantle rock, and, thereby, the likelihood of plate tectonics (Spaargaren et al. 2020). Within the typical range of major-element variability across the galaxy, viscosity can easily vary by 2–3 orders of magnitude.

The degree of differentiation of iron between the mantle and the core also influences mantle physical properties. Mantle iron content generally increases with oxygen abundance (high oxidation state of mantle minerals). An increased iron content promotes melting and volcanism, as well as reducing mantle viscosity, with important effects on plate tectonic activity at the surface. Iron can also induce mantle layering (Spaargaren et al. 2020), thereby sharply decreasing convective vigor and promoting a stagnant lid at the surface. However, for increasingly exotic exoplanetary compositions—those far from well-studied Earth (or Mars) rocks—our understanding of the relevant mantle properties becomes more and more unreliable.

It is, therefore, essential to design new experiments and models to explore the coupled interior–surface evolution of rocky planets as a function of planet mass, core size, mantle viscosity, surface temperatures, internal heat sources, and other factors. While the variety of exoplanetary conditions makes this endeavor very challenging, realistic and coupled interior–exterior models are required to link any upcoming observations of exoplanetary surfaces and atmospheres with the underlying interior dynamics. In this endeavor, our decent knowledge of the system Earth and of other terrestrial planets in the Solar System needs to be applied to the onrush of data from new exoplanet observation missions. As the planetary interior–atmosphere system is tightly coupled, observations of atmospheric properties may provide constraints on the surface tectonic regime of exoplanets, on the style of mantle convection, or even directly on the abundance of life itself.

## REFERENCES

Davaille A, Smrekar SE, Tomlinson S (2017) Experimental and observational evidence for plume-induced subduction on Venus. Nature Geoscience 10: 349-355, doi: 10.1038/ngeo2928

Dehant V and 17 coauthors (2019) Geoscience for understanding habitability in the solar system and beyond. Space Science Reviews 215: 42, doi: 10.1007/s11214-019-0608-8

Dorn C, Noack L, Rozel AB (2018) Outgassing on stagnant-lid super-Earths. Astronomy & Astrophysics 614: A18, doi: 10.1051/0004-6361/201731513

Foley BJ, Smye AJ (2018) Carbon cycling and habitability of Earth-sized stagnant lid planets. Astrobiology 18: 873-896, doi: 10.1089/ast.2017.1695

Forsyth D, Uyeda S (1975) On the relative importance of the driving forces of plate motion. Geophysical Journal International 43: 163-200, doi: 10.1111/j.1365-246X.1975.tb00631.x

Fulton BJ and 12 coauthors (2017) The California-*Kepler* survey. III. A gap in the radius distribution of small planets. Astronomical Journal 154: 109, doi: 10.3847/1538-3881/aa80eb

Greaves JS and 18 coauthors (2020) Phosphine gas in the cloud decks of Venus. Nature Astronomy doi: 10.1038/s41550-020-1174-4

Hamano K, Abe Y, Genda H (2013) Emergence of two types of terrestrial planet on solidification of magma ocean. Nature 497: 607-610, doi: 10.1038/nature12163

Haq BU, Schutter SR (2008) A chronology of Paleozoic sea-level changes. Science 322: 64-68, doi: 10.1126/science.1161648

Ingersoll AP (1969) The runaway greenhouse: a history of water on Venus. Journal of the Atmospheric Sciences 26: 1191-1198, doi: 10.1175/1520-0469(1969)026<1191:TRGAHO>2.0.CO;2

Kalousová K, Sotin C, Choblet G, Tobie G, Grasset O (2018) Two-phase convection in Ganymede's high-pressure ice layer — implications for its geological evolution. Icarus 299: 133-147, doi: 10.1016/j.icarus.2017.07.018

Kasting JF, Whitmore DP, Reynolds RT (1993) Habitable zones around main sequence stars. Icarus 101: 108-128, doi: 10.1006/icar.1993.1010

Kirschvink JL (1992) Late Proterozoic low-latitude global glaciation: the snowball Earth. In: Schopf JW, Klein C, Des Maris D (eds) The Proterozoic Biosphere: A Multidisciplinary Study. Cambridge University Press, New York, pp 51-52

Kite ES, Manga M, Gaidos E (2009) Geodynamics and rate of volcanism on massive Earth-like planets. Astrophysical Journal 700: 1732, doi: 10.1088/0004-637X/700/2/1732

Korenaga J (2013) Initiation and evolution of plate tectonics on Earth: theories and observations. Annual Review of Earth and Planetary Sciences 41: 117-151, doi: annurev-earth-050212-124208

Landuyt W, Bercovici D (2009) Variations in planetary convection via the effect of climate on damage. Earth and Planetary Science Letters 277: 29-37, doi: 10.1016/j.epsl.2008.09.034

Lourenço DL, Rozel AB, Ballmer MD, Tackley PJ (2020) Plutonic-squishy lid: a new global tectonic regime generated by intrusive magmatism on Earth-like planets. Geochemistry, Geophysics, Geosystems 21: e2019GC008756, doi: 10.1029/2019GC008756

Moore WB, Simon JI, Webb AAG (2017) Heat-pipe planets. Earth and Planetary Science Letters 474: 13-19, doi: 10.1016/j.epsl.2017.06.015

Noack L, Breuer D (2014) Plate tectonics on rocky exoplanets: influence of initial conditions and mantle rheology. Planetary and Space Science 98: 41-49, doi: 10.1016/j.pss.2013.06.020

Noack L and 6 coauthors (2014) Can the interior structure influence the habitability of a rocky planet? Planetary and Space Science 98: 14-29, doi: 10.1016/j.pss.2014.01.003

Noack L and 8 coauthors (2016) Water-rich planets: how habitable is a water layer deeper than on Earth? Icarus 277: 215-236, doi: 10.1016/j.icarus.2016.05.009

Roberts PH, Glatzmaier GA (2000) Geodynamo theory and simulations. Reviews of Modern Physics 72: 1081, doi: 10.1103/RevModPhys.72.1081

Rüpke LH, Morgan JP, Hort M, Connolly JAD (2004) Serpentine and the subduction zone water cycle. Earth and Planetary Science Letters 223: 17-34, doi: 10.1016/j.epsl.2004.04.018

Spaargaren RJ, Ballmer MD, Bower DJ, Dorn C, Tackley PJ (2020) The influence of bulk composition on long-term interior-atmosphere evolution of terrestrial exoplanets. Astronomy & Astrophysics 643: A44, doi: 10.1051/0004-6361/202037632

Stamenković V, Breuer D (2014) The tectonic mode of rocky planets: Part 1 – Driving factors, models & parameters. Icarus 234: 174-193, doi: 10.1016/j.icarus.2014.01.042

van Heck HJ, Tackley PJ (2008) Planforms of self-consistently generated plates in 3D spherical geometry. Geophysical Research Letters 35, doi: 10.1029/2008GL035190

Ward PD, Brownlee D (2003) Rare Earth: Why Complex Life is Uncommon in the Universe. Copernicus Books, New York, 335 pp

West AJ, Galy A, Bickle M (2005) Tectonic and climatic controls on silicate weathering. Earth and Planetary Science Letters 235: 211-228, doi: 10.1016/j.epsl.2005.03.020

## FIGURES

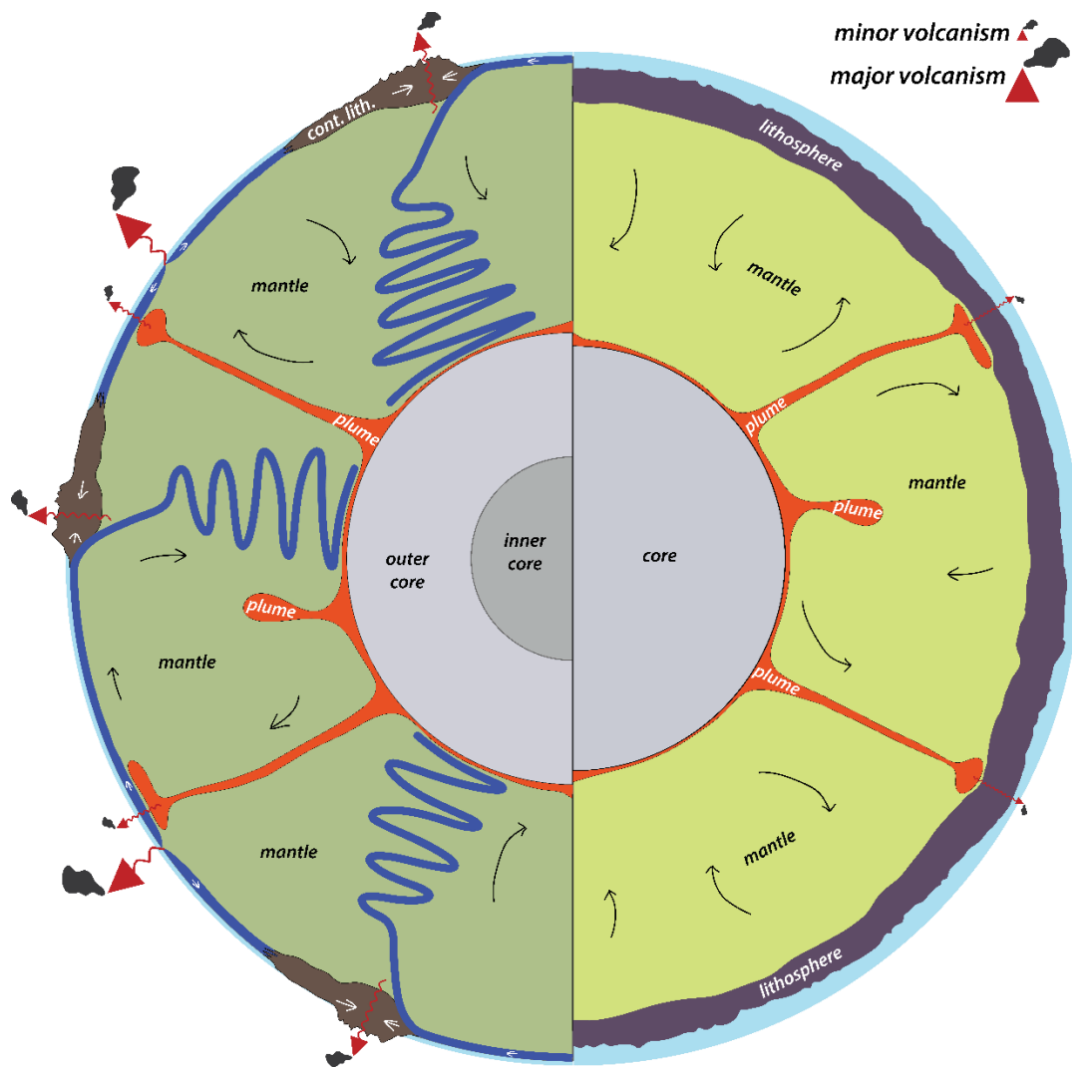

**FIGURE 1 (LEFT)** Dynamics of a plate-tectonic planet. **(RIGHT)** Dynamics of a stagnant-lid planet. Respective halves of composite planet not to scale. Continental lithosphere in dark brown; oceanic lithosphere in dark blue; stagnant lid in purplish brown; mantle in green. White arrows denote plate motion; black arrows denote mantle convection patterns. Hot upwelling plumes are orange. Triangles mark sites of more-or-less vigorous volcanism. Liquid outer core is light grey; solid inner core is dark grey. Potential oceans in light blue. The size of each planetary hemisphere is arbitrary.

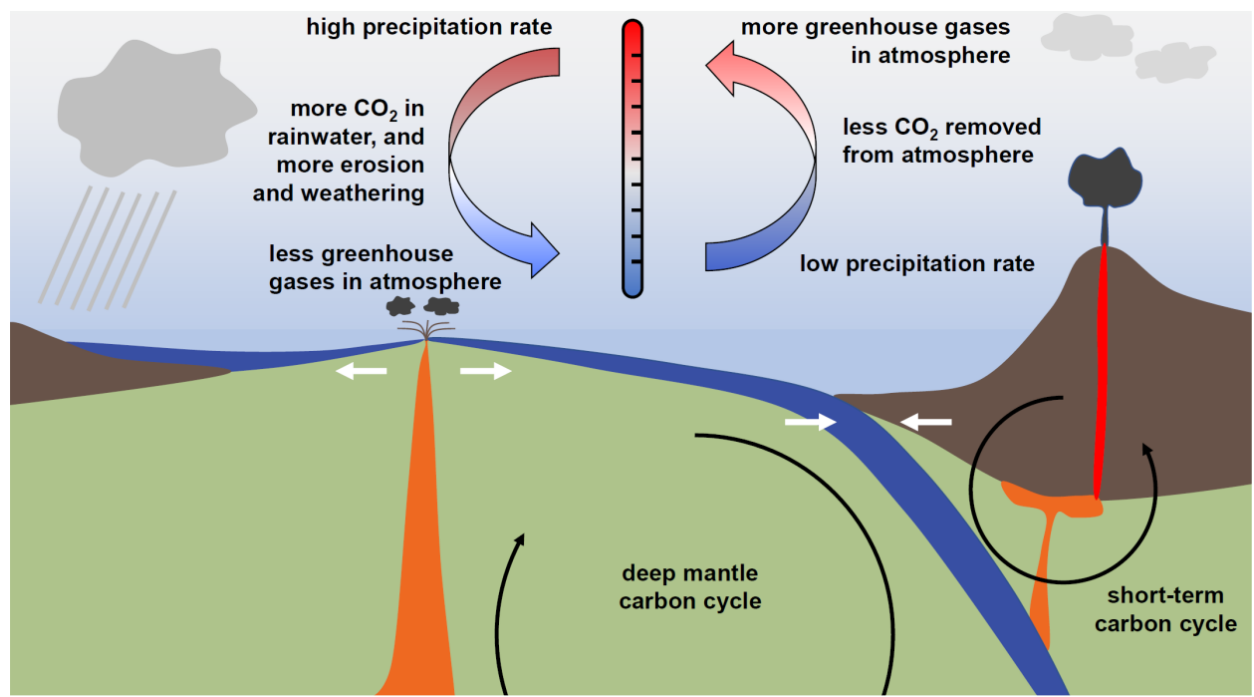

**FIGURE 2** Volatile cycles and buffering of climate due to silicate weathering on a plate tectonic planet.

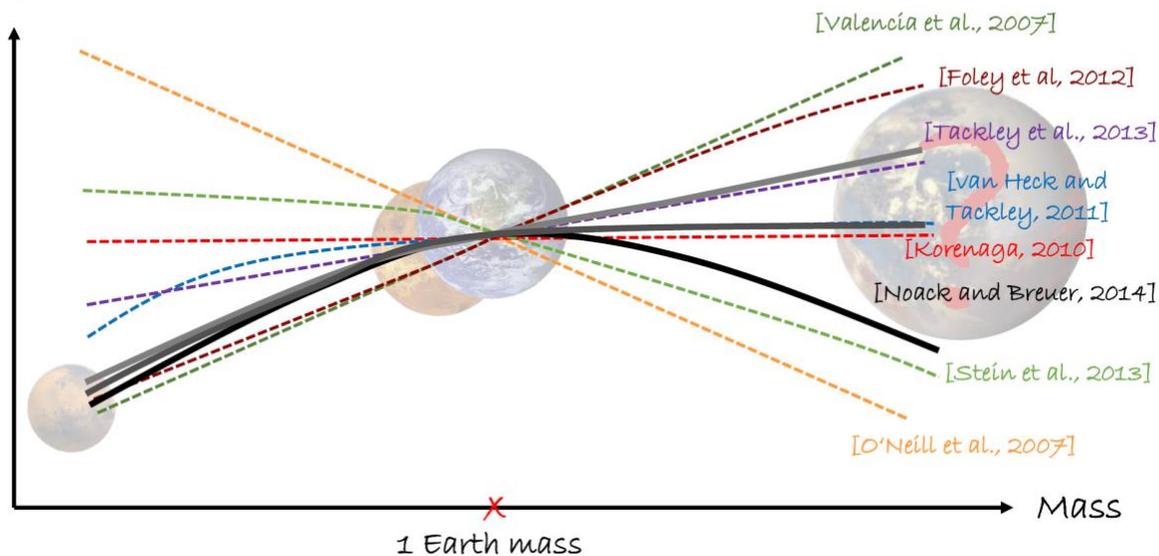

**FIGURE 3** Sketch summarizing different predictions for the likelihood of plate tectonics as a function of planet mass, scaled to the same likelihood at one Earth mass (as a reference point). The grey-to-black lines indicate different trends as a function of initial mantle temperature (grey = hot; black = cold) after planet formation as found in Noack and Breuer (2014). The dashed colored lines refer to previous studies as cited in Noack and Breuer (2014).